%% file: main.tex
\documentclass{wscpaperproc}
\usepackage{latexsym}
\usepackage{graphicx}
\usepackage{mathptmx}
\usepackage[T1]{fontenc}

\usepackage{amsmath}
\usepackage{amsfonts}
\usepackage{amssymb}
\usepackage{amsbsy}
\usepackage{amsthm}
\usepackage{tabularx}
\usepackage{subfig}

\usepackage[pdftex,colorlinks=true,urlcolor=blue,citecolor=black,anchorcolor=black,linkcolor=black]{hyperref}

\newtheoremstyle{wsc}
{3pt}
{3pt}
{}
{}
{\bf}
{}
{.5em}
{}

\theoremstyle{wsc}

\begin{document}

%
%

\pagestyle{fancyplain}

\thispagestyle{plain}
\firstPageHead{}

\chead{\fancyplain{}{\itshape Zhu, Vyetrenko, Grundl, Byrd, Dwarakanath, and Balch}}

\rhead{}
\cfoot{}
\renewcommand{\headrulewidth}{0pt} 

\input{wscbib.tex}           

\setlength{\baselineskip}{12.7pt}

\title{ONCE BURNED, TWICE SHY? THE EFFECT OF STOCK MARKET BUBBLES ON TRADERS THAT LEARN BY EXPERIENCE}

\author{
    Haibei Zhu\\
    Svitlana Vyetrenko\\
    Kshama Dwarakanath\\
    Tucker Balch\\[12pt]
    J.P. Morgan AI Research\\
    383 Madison Avenue\\
    New York, NY 10017, USA\\
    \and
    Serafin Grundl\\ \\ \\ \\[12pt]
    Federal Reserve Board\\
    Constitution Avenue NW \&, 20th Street NW\\
    Washington, DC 20551, USA\\
    \and
    David Byrd\\[12pt]
    Bowdoin College\\
    233 Main Street\\
    Brunswick, ME 04011, USA\\
}

\maketitle

\section*{ABSTRACT}
We study how experience with asset price bubbles changes the trading strategies of reinforcement learning (RL) traders and ask whether the change in trading strategies helps to prevent future bubbles. We train the RL traders in a multi-agent market simulation platform, ABIDES, and compare the strategies of traders trained with and without bubble experience. We find that RL traders without bubble experience behave like short-term momentum traders, whereas traders with bubble experience behave like value traders. Therefore, RL traders without bubble experience amplify bubbles, whereas RL traders with bubble experience tend to suppress and sometimes prevent them. This finding suggests that learning from experience is a mechanism for a boom and bust cycle where the experience of a collapsing bubble makes future bubbles less likely for a period of time until the memory fades and bubbles become more likely to form again.

\section{Introduction}
\label{ch_introduction}

Financial bubbles are a well-known economic phenomenon defined as periods of unsustainable growth or loss in the price of an asset \shortcite{sornette2014financial}. In a bubble, asset prices deviate from their intrinsic valuation \shortcite{garber1990famous}. The occurrence of financial bubbles accompanies the increase of trade in all realms \shortcite{frehen2013new}. Bubbles are dangerous and can significantly damage the existing economic system \shortcite{ofek2003dotcom,phillips2011dating}. While the mechanism of how a bubble generates and bursts has been well studied \shortcite{phillips2018financial}, few efforts have focused on how such bubbles may affect investor trading strategies and how investors can potentially affect the evolution of existing bubbles. In this work, we investigated how exposure of market agents to bubble scenarios in a simulated environment affected future trading strategies and bubble formation. We found that market participants with sufficient bubble experience tend to suppress bubble formation and lessen bubble duration while maintaining profitability.

Our empirical results derive from a financial market multi-agent system (MAS) based on the Agent-Based Interactive Discrete Event Simulation (ABIDES) environment \shortcite{byrd2019abides,vyetrenko2020get}. By configuring each agent as an individual market participant, MAS can effectively represent a financial market \shortcite{lebaron2001builder}. In ABIDES, agents can communicate only via messages routed through the simulation kernel with appropriate computation and communication delays. Similar to the real world, most of these will be order-related activity and status messages to or from a stock exchange agent. We populate the simulation with market participants following various pre-defined trading strategies and configure it to generate market bubbles. Our experimental agents utilize reinforcement learning (RL) to learn from the experience that they are exposed to in order to optimize a trading policy that should earn profit and potentially affect market bubbles. We train RL agents with different quantities of bubble exposure as measured by the percentage of training scenarios that contain a bubble. We then evaluate these learning-based agents on new bubble scenarios to investigate their tendency to enhance or reduce bubbles' formation, size, and duration.

The main contribution of this work is a detailed simulated study of how experiencing stock market bubbles ``trains'' the learning market agents to modify their trading strategies from momentum-based to value-based on the purpose of maximizing their profits -- which also results in fewer and smaller bubbles. This paper is organized as follows. Section \ref{ch_background} covers motivation and related work. Section \ref{ch_experiment} describes the simulation experiment configurations, followed by the experiment results in Section \ref{ch_results}. We discuss the results in Section \ref{ch_discussion} and conclude our work with future directions in Section \ref{ch_conclusion}.

\section{Background}
\label{ch_background}

The history of financial bubbles and the mechanism of how such bubbles form and burst have been well investigated in the literature \shortcite{brunnermeier2013bubbles,bhattacharya2008causes}. Specifically, the life cycle of a bubble has been clearly illustrated \shortcite{scherbina2014asset}. A bubble usually starts with a new expectation of an asset or a random price fluctuation. For instance, in an upward bubble, buy orders flow in and significantly increase the price because potential payoff attracts investors. Such a rapid price increase and potential return attract further buy orders. However, fast price inflation is usually followed by a quick decrease -- the burst of a bubble. A bubble can be burst for various reasons, including regulatory constraints \shortcite{lyocsa2022yolo} and liquidity decrease \shortcite{sornette2014financial}. When bubbles burst, the equilibrium in the order book breaks, and the sell orders dominate.

The primary focus of this work is the effect of prior bubble experience on trading strategies. Human-subject experiments introduce variations that can be confounding factors. Our approach relies on a multi-agent market simulation to better isolate the effect of the bubble experience. Multi-agent systems have been widely used in financial studies \shortcite{lux1999scaling,preis2006multi,byrd2019explaining}. Compared to human-subject experiments, multi-agent systems can simulate markets with a large scale of market participants in an infinite time horizon, which cannot be achieved in lab studies. Market participants in simulations can be assigned to various empirically designed trading strategies. Also, such simulations allow researchers to control market variables precisely via specified configurations to answer specific research questions \shortcite{balch2019evaluate}.

This study used ABIDES as the multi-agent platform for simulating the financial market \shortcite{byrd2019abides,amrouni2021abides}. The core of ABIDES is a simulation kernel with a messaging system that allows agents to communicate with configurable latency and nanosecond resolution in continuous double-auction trading similar to NASDAQ. Each stock offered in ABIDES is configured with the help of an exogenous time series that represents the agent's understanding of the outside world -- the ``fundamental'' price \shortcite{byrd2019explaining}, which is a mean-reverting time series based on the Ornstein-Uhlenbeck (O-U) process \shortcite{uhlenbeck1930theory}.

Some efforts have been made to simulate financial bubbles using multi-agent systems \shortcite{duffy2006asset,samanidou2007agent,westphal2020market}. However, these studies have a limitation in configuring experimental agents with rule-based trading strategies. Agents follow one or a few strategies throughout the simulation, and such strategies are not guaranteed to adapt to market changes and obtain payoffs. Understanding that traders in the real world have more complicated strategies, which monitor market environment states, including holding positions, volumes, and trends, learning-based agents are needed to better approximate real-world trading strategies \shortcite{kolm2020modern,charpentier2021reinforcement}.

\begin{figure}
    \centering
    \includegraphics[width=\columnwidth, height=0.34\textheight]{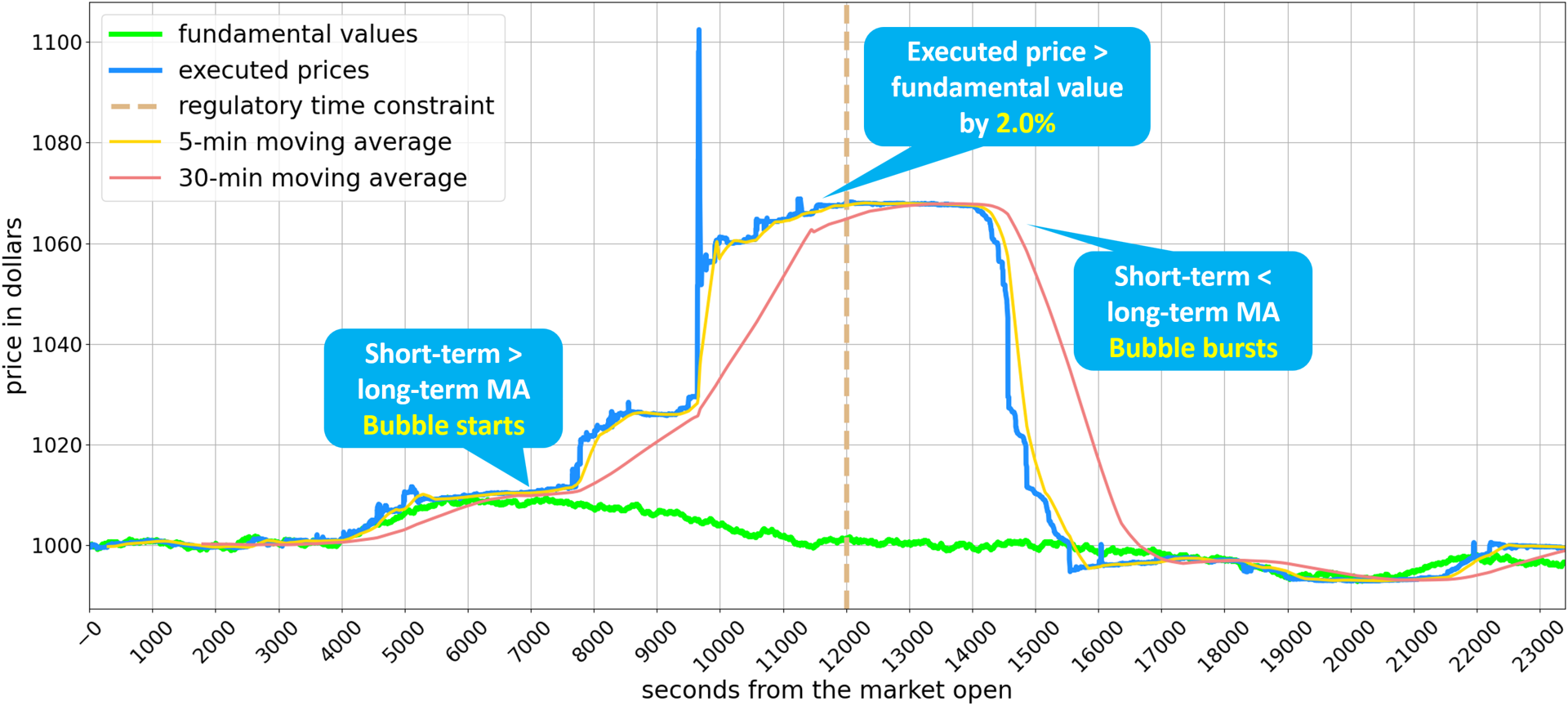}
    \caption{The start and end of a bubble are identified by comparison of moving averages on the market price with the ``fundamental'' value of the asset.}
    \label{fig_definition_bubble}
\end{figure}

\section{Experiment}
\label{ch_experiment}

We designed and conducted the experiments in ABIDES. We first configured an exchange agent to process all messages -- buy or sell orders placed by other agents. Once the exchange agent receives the orders, it matches them within the order book and executes them. The exchange agent also publishes the order book's status and the executed orders' details so that other agents can observe how the stock price changes and how the order book evolves. Our study only focused on one stock to simplify the market environment. Since shocks and external impacts are not considered, we configured a mean-reverting fundamental time series as the stock's expected valuation. A market maker is also configured as a special agent, which provides market liquidity by placing orders on both sides of the order book. Specifically, we configured four types of rule-based background agents with corresponding strategies:

\begin{itemize}
    \item \textbf{Value agents} represent professional investors with a better understanding of stock valuation. This is the only role with access to the fundamental value of a stock. The agent places orders to arbitrage market prices against the valuation, which is an approximate observation of the fundamental value -- buying if the stock is undervalued and selling if it is overvalued.
    \item \textbf{Noise agents} represent retail investors who place a few orders per trading day. They follow a simple strategy that places orders in a random direction at random times. Unlike professional investors, retail investors have a limited understanding of the stock valuation and the market trend. Also, the order volume from retail investors is usually less than professional investors.
    \item \textbf{Momentum agents} follow a simple but aggressive momentum-oriented strategy. The momentum is determined by comparing the short-term (5-min) and long-term (30-min) moving averages of the stock price. Momentum agents place buy orders when the short-term exceeds the long-term moving average price. Their order quantities are similar to value agents. Momentum agents utilize limit orders, which include a minimum price the agent will accept to sell or a maximum price it will pay to buy. Due to this constraint, limit orders are not guaranteed to be executed.
    \item \textbf{Herding agents} are similar to momentum agents, monitoring price momentum and placing orders accordingly. But herding agents place market orders, which guarantee immediate execution at the best available price. So, they can rapidly push the executed price upward or downward.
\end{itemize}

\subsection{Generating and Bursting Bubbles}

As mentioned in Section \ref{ch_background}, the price deviation at the early stage of a bubble starts with overwhelming orders in one direction. Given this fact, we created bubble scenarios by configuring momentum and herding agents -- both can place a large number of orders in the same direction within a short amount of time. Specifically, we included 1) 50 value agents, 2) 500 noise agents, 3) 8 momentum agents, 4) 5 herding agents as rule-based market participants, and 6) a market maker agent in the simulations to create bubble scenarios. The percentage of different types of agents is similar to the real market. In addition to generating bubbles, we created non-bubble scenarios by removing herding agents from the simulations. An example of creating a bubble in the simulations is shown in Figure \ref{fig_definition_bubble}. The green time series represents the fundamental value of a stock, and the blue curve represents the executed price. Due to random price fluctuation, the short-term moving average exceeded the long-term value at around the timestamp of 10000 seconds. Herding agents then placed a considerable amount of buy orders that significantly increased the executed price.

Understanding that herding agents share an aggressive trading strategy and can break the balance of the order book rapidly, we burst bubbles by setting a time constraint on herding agents that they cannot place any orders after 12000 seconds when the intraday market simulations start. Such a time constraint mimics regulators' action to reduce market volatility and protect the market. This time constraint is also illustrated in Figure \ref{fig_definition_bubble} that after 12000 seconds, herding agents stopped placing buy orders and sell orders from value agents dominate the order book to push the price to converge back to the fundamental value. Thus, the executed price shown in blue demonstrates an example of the generation and burst of a bubble in our simulations.

\subsection{Bubble Detection}

Previous studies have proposed methods for determining the start and burst of bubbles \shortcite{jarrow2011detect,shu2020real,phillips2019detecting}. Since we only simulate one stock without external impacts, we follow an intuitive metric to detect bubbles, including two measures -- 1) comparing the short-term and long-term moving averages of the stock price, and 2) comparing the executed price and the fundamental value. It is worth noting that the second measure would only work in a simulated environment. Because in real markets, the fundamental price is assumed but not explicitly available.

Figure \ref{fig_definition_bubble} also illustrates an example of the two measures in the bubble detection metric. The yellow and brown curves are the short-term and long-term price moving averages. For a bubble with upward deviated prices, as shown in Figure \ref{fig_definition_bubble}, we consider the time point when the short-term exceeds the long-term moving average as the starting of a bubble. The ending of a bubble is when the short-term average value is lower than the long-term average. The main body of the bubble also needs to meet the second measure that the executed price should be larger than the fundamental price by 2.0\%.

\begin{table}[htb]
    \caption{Reinforcement learning agent state features.}
    \label{table_rl_agent_states}
    \renewcommand{\arraystretch}{1.0}
    \centering
    \begin{tabularx}{\textwidth}{c|c|X}
        \hline
        Index & Name & Description \\
        \hline
        1 & Holding & The agent's current holding \\
        2 & Imbalance & The difference between the best buy and sell order volume \\
        3 & Volatility & The standard deviation of a 30-min price history \\
        4 & Mid-price & The latest executed price or the average of the best buy and sell price \\
        5-9 & Momentum x-min & The x-min (x is 30, 60, 90, 120, or 180) momentum is an indicator of the comparison between the 5-min and x-min moving average of the price \\
        \hline
    \end{tabularx}
\end{table}

\begin{table}[htb]
    \caption{Bubble measurement metrics.}
    \label{table_bubble_metrics}
    \renewcommand{\arraystretch}{1.0}
    \centering
    \begin{tabularx}{\textwidth}{c|c|X}
        \hline
        Bubble & Metrics & Method / Calculation \\
        \hline
        Detection & If is a bubble & 1) Comparison of moving averages; 2) Fundamental value as a reference\\
        \hline
        & Count & If is a bubble, then count increments by 1 \\
        Measurement & Magnitude & The average difference between the fundamental value and the mid-price in a bubble \\
        & Duration & The average elapsed time in a bubble \\
        \hline
    \end{tabularx}
\end{table}

\subsection{Reinforcement Learning Agent Training and Testing}

Reinforcement Learning (RL) is a tool for sequential decision-making that allows learning trading agents that execute optimally given the simulated environment. Like rule-based agents, RL agents also perceive the market environment and place orders accordingly in the order book. But unlike rule-based agents, they are expected to adapt to various market situations and are trained to have the optimal capability of earning profits. The RL agents perceive and interpret the market through the state space, which includes nine features explained and shown in Table \ref{table_rl_agent_states}. These features were designed and configured according to the status of the RL agent (like the ``holding" feature) and the common indicators for the stock market (like ``imbalance", ``volatility", ``mid-price", and ``momentum"). The RL agent's policy network takes the feature values as inputs and outputs actions to maximize the agent's reward. Therefore, we can model trading decisions based on signals derived from market knowledge by a Markov Decision Process and train reinforcement learning trading agents in ABIDES.

RL agents can conduct three simulation actions: 1) buy, 2) hold or do nothing, and 3) sell. Both buy and sell actions place market orders with a fixed quantity of shares throughout the experiments. The reward for the RL agent is the instant profit, which is the marked-to-market value difference between the current and the previous timestamp. The marked-to-market value is the total value of the remaining cash and the current holding volume times the latest stock mid-price. In other words, RL agents do not need to clean holding positions to realize profits. RL agents will be activated every minute in simulations and complete a cycle of observing the market status, determining and conducting an action, and receiving a reward.

RL agents were trained in five market environments with different levels of bubble experience. Such levels were represented in the increasing percentages of bubble scenarios among all training scenarios, including 0\% (without bubble experience), 25\%, 50\%, 75\%, and 100\% (full bubble experience). Agents share the same state space, action space, and reward function while training. The Proximal Policy Optimization (PPO) algorithm was utilized for training these agents \shortcite{schulman2017proximal}. The first RL agent was trained with only non-bubble scenarios. Other RL agents were trained with bubble scenarios. The percentage of bubble scenarios during training was controlled by the random seeds in configuring agents and the market environment. In bubble scenarios, we applied the specific random seeds generated beforehand to ensure the scenarios included at least one bubble.

Once we had trained RL agents, they were applied to market simulations to evaluate their profit performance and to observe their behaviors in bubble scenarios. Our main interest in the testing sessions is the agents' capability to affect the evolution of bubbles. Thus, we defined bubble measurement metrics in which three measures were observed and calculated. The metrics include bubble 1) count, 2) magnitude, and 3) duration, as illustrated in Table \ref{table_bubble_metrics}.

\section{Results}
\label{ch_results}

RL agents trained with different levels of bubble experience were applied to the testing sessions with 100\% bubble scenarios respectively to investigate agents' performance in terms of profit and affecting bubbles. We applied each RL agent to the simulation with another set of pre-defined bubble seeds for 1000 runs and collected experiment results.

\begin{figure}
    \centering
    \subfloat[Resulting profit distribution comparison.]{
        \label{fig_profit_analysis}
        \includegraphics[width=0.49\columnwidth, height=0.24\textheight]{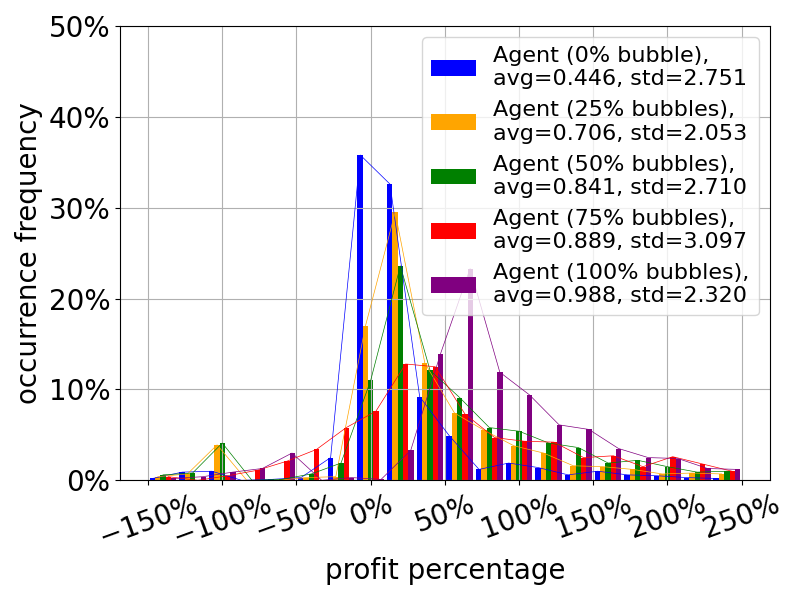}}
    \hfil
    \subfloat[Resulting bubble count distribution comparison.]{
        \label{fig_bubble_count}
        \includegraphics[width=0.49\columnwidth, height=0.24\textheight]{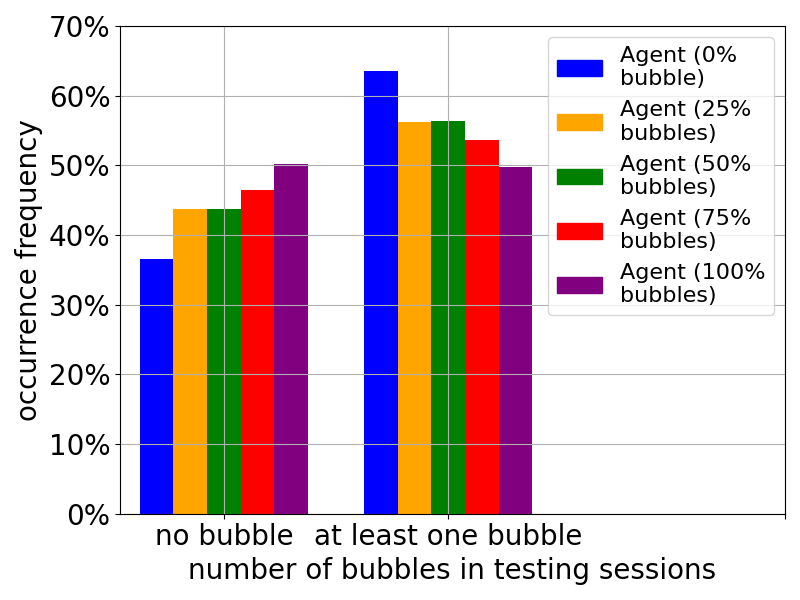}}
    \caption{Profit analysis and bubble count testing results.}
\end{figure}

\subsection{Profit Analysis}

The first experimental outcome is the profit analyses on RL agents trained with different bubble conditions. Such profit analyses illustrate RL agents' profit capability in markets with bubbles. The histogram in Figure \ref{fig_profit_analysis} presents the resulting profit distribution for the five RL agents. The RL agent trained without bubble experience has an average profit of 0.446, which represents 44.6\% of the starting cash. Meanwhile, the standard deviation of the profit is 2.751, which indicates that the agent may result in a huge gain or loss due to the high market volatility in some scenarios. The agent trained with 25\% bubble experience has a slightly higher profit average of 70.6\%. Similarly, it also has a large standard deviation of 2.053.

As we expected, agents trained with a higher percentage of bubble experience have a higher expectation of profit -- the agent trained with 100\% bubble experience has an average profit of 98.8\%. While these five profit distributions can be distinguished clearly due to the significant difference in averages, interestingly, the profit standard deviations of these agents are similar. This indicates that even agents trained with a higher percentage of bubble experience can be affected by the large market volatility in bubble scenarios and can hardly maintain a stable profit level.

\subsection{Bubble Measurement Metrics}

The main results for answering our research questions are based on the three bubble measurement metrics mentioned in Table \ref{table_bubble_metrics}. The resulting measures from the five agents were plotted as multi-group histograms to illustrate how RL agents affect the bubble measurement metrics and the evolution of bubbles.

\subsubsection{Bubble Count}

The first metric is the bubble count. As shown on the left portion of Figure \ref{fig_bubble_count}, bubbles do not appear in over 35\% of the resulting runs for all agents' testing sessions. And a monotonic trend indicates that the higher percentage of bubble scenarios in the agent's training, the higher the percentage of testing runs with no bubble. Another clear trend on the right portion of Figure \ref{fig_bubble_count} shows that for the testing scenarios containing bubbles, the average number of bubbles decreases with the increase of agent bubble experience. Such a fact supports that agents with more bubble experience can suppress bubbles better.

A non-parametric statistical test of the Kruskal–Wallis one-way analysis of variance with a significance level of $\alpha =0.05$ was conducted on the bubble count measures. Significance was observed that the p-value is less than 0.001 and the $\alpha$. This indicates that the level of bubble experience in RL agent training significantly impacts bubble count in testing sessions -- agents with more bubble experience during training are better able to reduce the number of bubbles generated. Since agents' only reward is profit, agents that have experienced bubbles would trade against the trends of bubbles to ensure profits. The more bubble scenarios an agent experienced, the stronger the agent would counter the bubbles.

\begin{figure}
    \centering
    \subfloat[Resulting bubble magnitude distribution comparison.]{
        \label{fig_bubble_magnitude}
        \includegraphics[width=0.49\columnwidth, height=0.24\textheight]{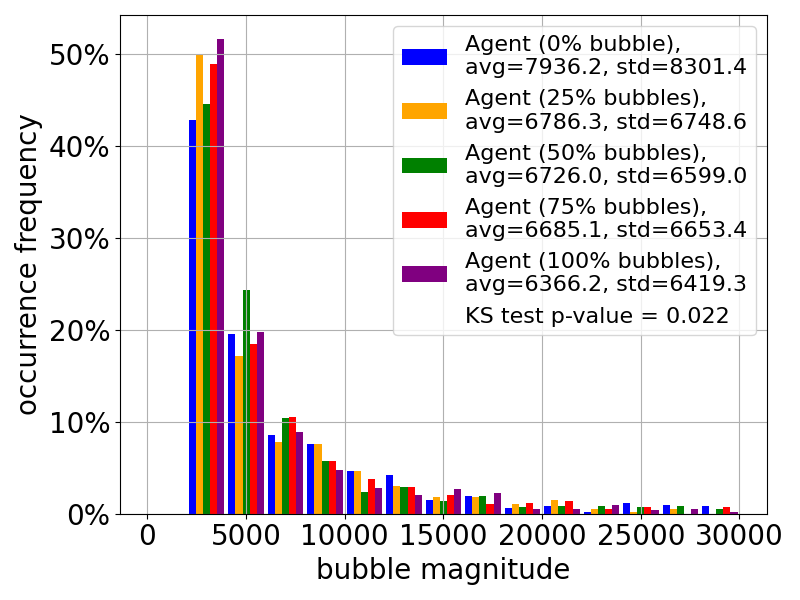}}
    \hfil
    \subfloat[Resulting bubble duration distribution comparison.]{
        \label{fig_bubble_duration}
        \includegraphics[width=0.49\columnwidth, height=0.24\textheight]{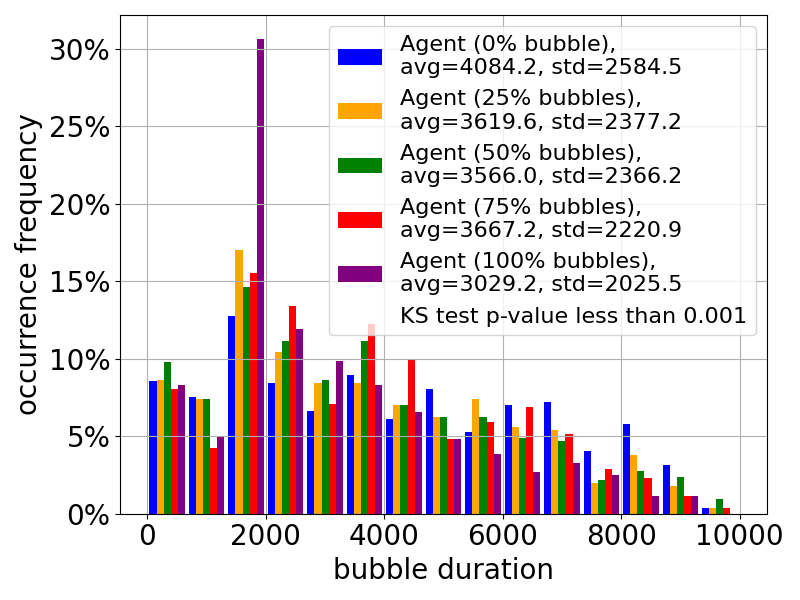}}
    \caption{Bubble magnitude and duration testing results.}
\end{figure}

\subsubsection{Bubble Magnitude}

The second metric is the bubble magnitude, the average difference between the executed price and the fundamental value within a bubble. As shown in Figure \ref{fig_bubble_magnitude}, the average resulting bubble magnitude for the RL agent trained without bubble experience is 7936.2 with a standard deviation of 8301.4. Similarly, with the increase in agent bubble experience, the resulting bubble magnitude decreases in the testing sessions. The average magnitude drops to 6366.2 with a standard deviation of 6419.3 for the agent trained with 100\% bubble experience. The Kruskal–Wallis test result also shows significance with a p-value of 0.022, smaller than the significance level of $\alpha =0.05$. Thus, the agent bubble experience significantly affects the resulting bubble magnitude. As with bubble counts, greater exposure to bubbles during training causes a learning agent to trade counter to them and increasingly suppress their magnitude.

\subsubsection{Bubble Duration}

The third metric is the bubble duration, which is the average elapsed time of bubbles in testing runs. The distribution of the resulting duration for all testing sessions is shown in Figure \ref{fig_bubble_duration}. The average bubble duration for the RL agent trained without bubble experience is 4084.2 seconds with a standard deviation of 2584.5. The average bubble duration for the agent trained with 100\% bubble experience is 3029.2 with a standard deviation of 2025.5. The Kruskal–Wallis test result supports the observation that the average duration slightly decreases if the agent trained with a higher percentage of bubble scenarios with a p-value less than 0.001, smaller than the significance level of $\alpha =0.05$. So, we can conclude that the level of bubble experience in agents' training impacts the resulting bubble duration during testing.

\begin{figure}
    \centering
    \subfloat[The RL agent trained with 0\% (without) bubble experience.]{
        \label{fig_bubble_example_mkt_0}
        \includegraphics[width=0.99\columnwidth, height=0.44\textheight]{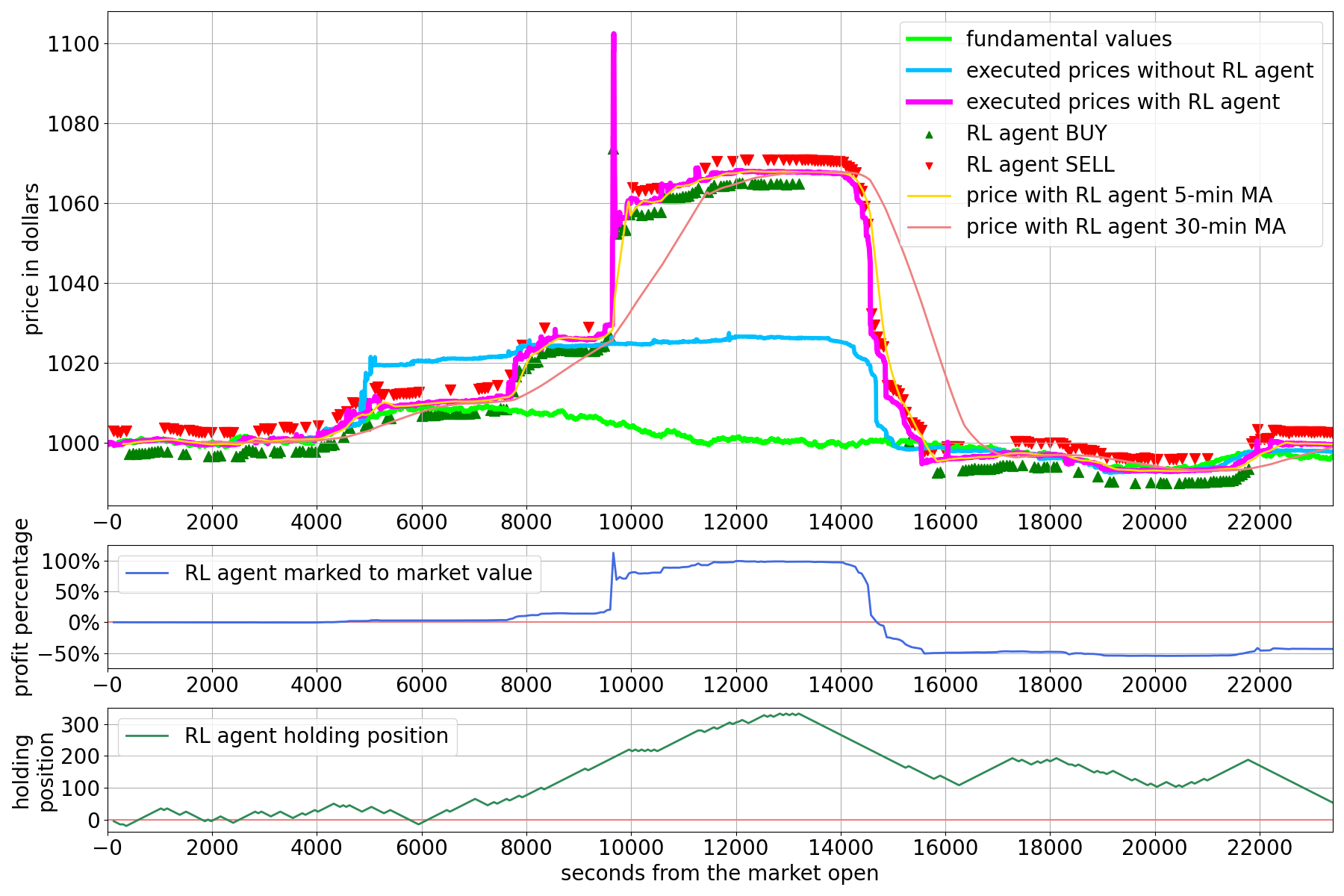}}
    \hfil
    \subfloat[The RL agent trained with 100\% bubble experience.]{
        \label{fig_bubble_example_mkt_100}
        \includegraphics[width=0.99\columnwidth, height=0.44\textheight]{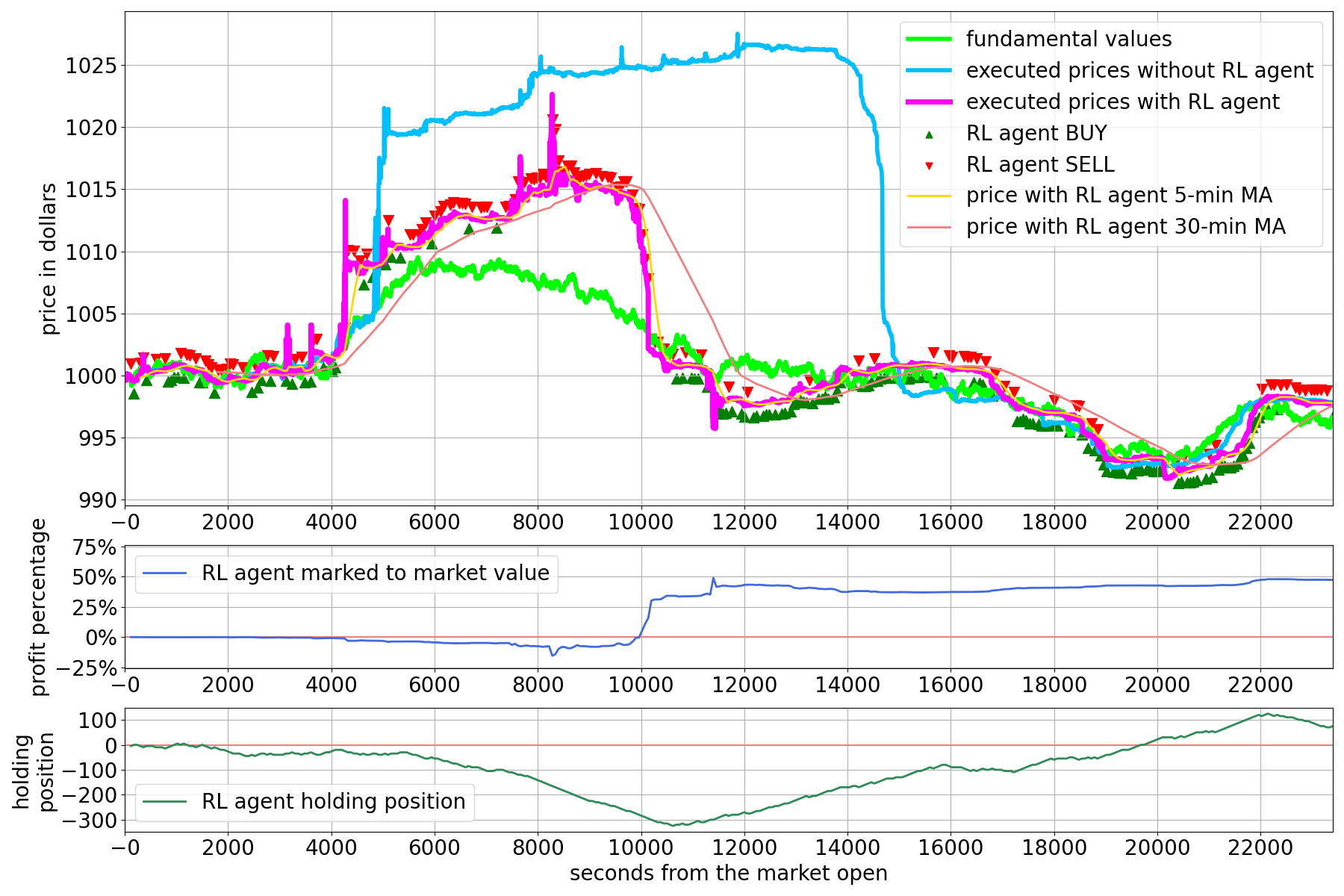}}
    \caption{A bubble scenario example with RL agent holding position and marked-to-market profit.}
    \label{fig_bubble_example}
\end{figure}

\begin{figure}[htb]
    \centering
    \subfloat[The agent trained with 0\% bubble experience.]{
        \label{fig_shap_mkt_0}
        \includegraphics[width=0.48\columnwidth, height=0.21\textheight]{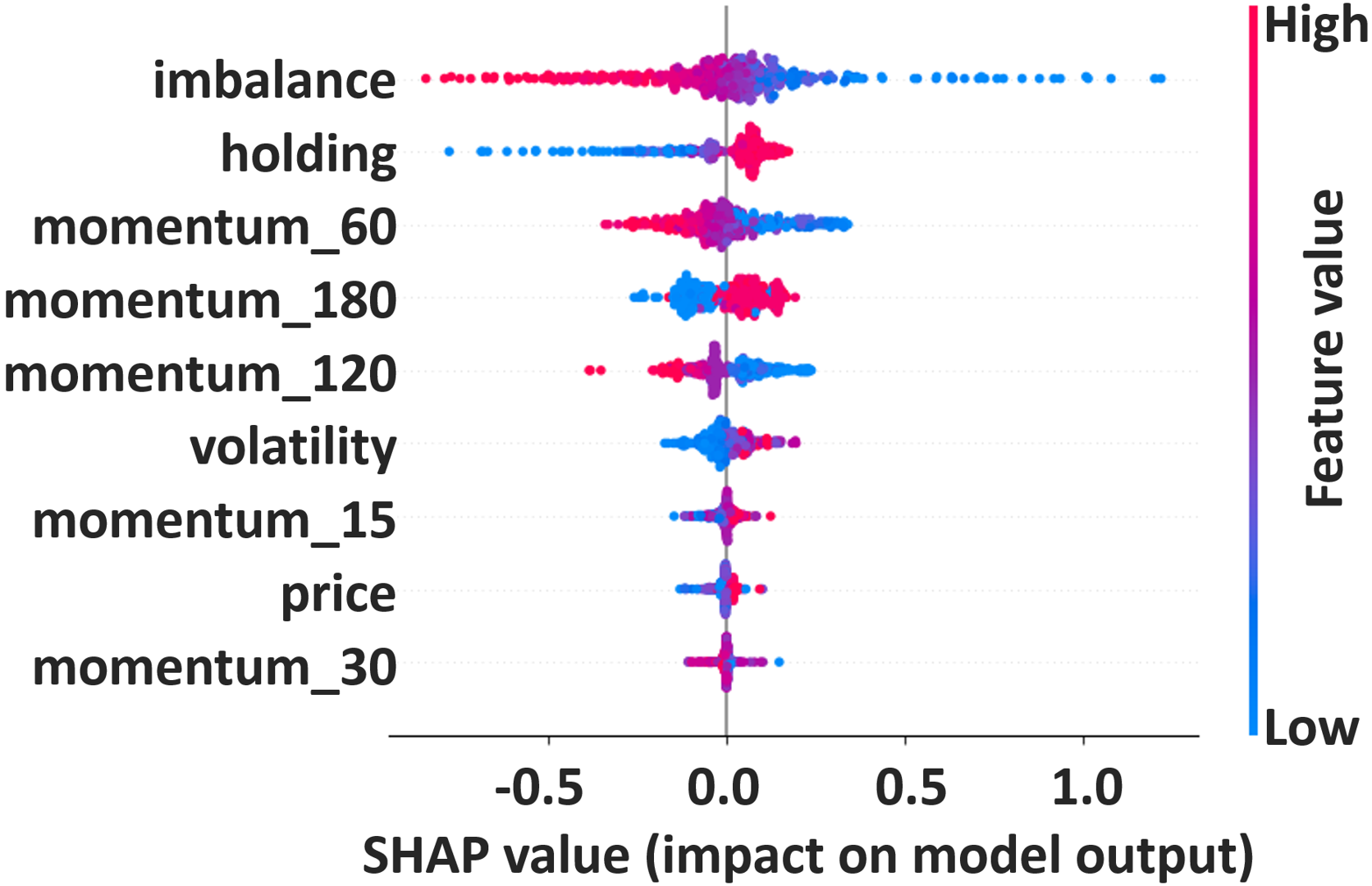}}
    \hfil
    \subfloat[The agent trained with 100\% bubble experience.]{
        \label{fig_shap_mkt_100}
        \includegraphics[width=0.48\columnwidth, height=0.21\textheight]{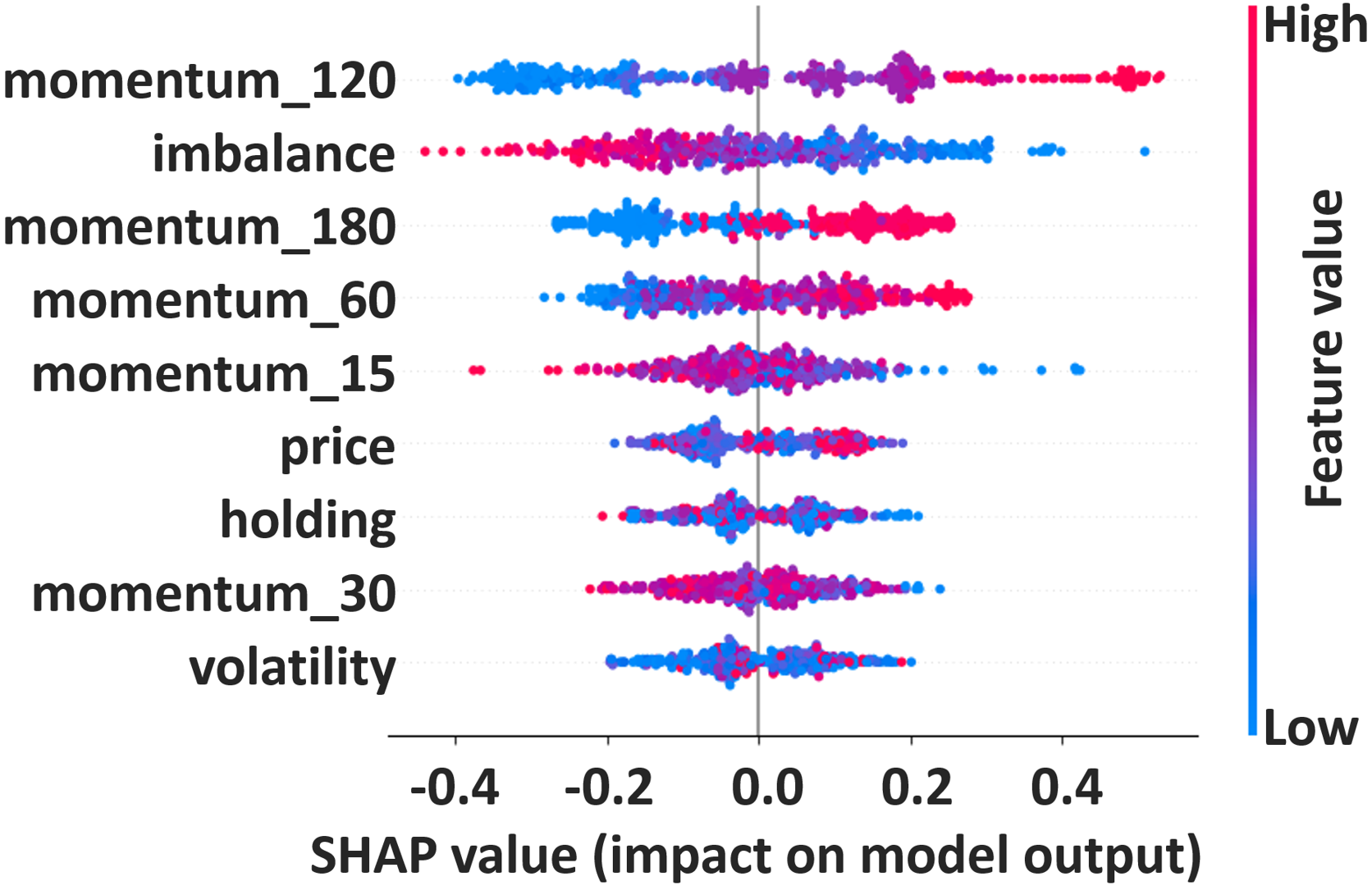}}
    \caption{Cumulative SHAP analysis for both agents in the example bubble scenario.}
    \label{fig_shap_analysis}
\end{figure}

\section{Discussion}
\label{ch_discussion}

The RL agent testing results shown in the previous section illustrate that the level of bubble experience in the agent training process would affect the agent's capability of countering and suppressing bubbles. An agent trained with a higher percentage of bubble scenarios would perform a trading strategy with a better capability of suppressing bubbles. So, it is anticipated that the performance and tactics of the agents trained with 0\% and 100\% bubble experience will vary significantly. To further analyze and compare agent strategies, we applied these two agents to a typical bubble scenario to investigate the bubble formation in the market and infer how the market environment would influence agent decisions and actions.

The resulting markets with RL agents' marked-to-market profit and holding position were plotted as Figure \ref{fig_bubble_example_mkt_0} and \ref{fig_bubble_example_mkt_100}. The light blue curves in the main chart represent the executed prices generated from the simulation with the same market environment but without RL agents. A bubble appears in this scenario, starting at about 5000 seconds after the trading starts and ending at about 15000 seconds. At the bubble's peak, the executed price deviated from the fundamental value of over 2.5\%. The green curve indicates the mean-reverting fundamental value of the stock, representing the stock's intrinsic value. The fundamental value varies within a limited range of 1.0\%.

\subsection{Bubble Formation}

The resulting bubble formation in Figure \ref{fig_bubble_example_mkt_0} and \ref{fig_bubble_example_mkt_100} has a major difference illustrated by the executed prices with RL agents. Applying the RL agent trained without bubble experience, the bubble still exists in the market, as shown in pink. The resulting bubble's price deviation doubled compared to the original value. The green and red markers represent RL agents' ``buy'' and ``sell'' orders. While there were some ``sell'' orders, most orders placed by the agent were ``buy'' orders during the bubble generation period. This is also supported by the RL agent holding position plot in Figure \ref{fig_bubble_example_mkt_0} that the holding volume increased when the stock price surged. Such buy orders caused a huge loss to the agent at the end of the trading day since the agent could not sell all holdings within a short period when the bubble burst. The agent's profit is indicated in the marked-to-market value plot as the percentage of the starting cash.

The agent trained with 100\% bubble experience suppressed the original bubble's price deviation. The maximum deviation between the executed price and the fundamental value was reduced to less than 1.0\% of the fundamental value. Thus, no bubble appears in this case. At about 5000 seconds, the executed price increased due to the overwhelming ``buy'' orders from the momentum and herding agents. However, the RL agent placed ``sell'' orders to the opposite side of the biased order book. This is reflected in the holding position plot in Figure \ref{fig_bubble_example_mkt_100}. Such ``sell'' orders balanced the order book and suppressed the original bubble. Also, these ``sell'' orders brought a large profit to the RL agent after the executed price dropped.

\subsection{RL Agent Strategy Analyses}

We ran the SHAP analysis throughout the trading day. And we collected the corresponding SHAP values for the nice features in RL agents' state space at all timestamps to form cumulative SHAP value scatter plots shown in Figure \ref{fig_shap_analysis}. The cumulative SHAP value plots illustrate the overall feature importance ranking in the state space based on all conducted actions by the agents. Different scatter colors for a specific feature represent the degree of that feature contributing to agent actions.

Using Figure \ref{fig_shap_mkt_0}, which illustrates the feature importance for the agent trained without bubble experience, as an example, the red scatters of the ``imbalance'' feature represent a high imbalance in the order book, and such values lead to negative SHAP values, which indicate the ``buy'' action in RL agents' action space. In other words, if the ``buy'' orders or bids outweigh the ``sell'' orders or asks, the agent prefers to place ``buy'' orders and intensify the imbalance in the order book. This agent also relies heavily on its holding positions and the 60-min momentum of the stock price. As the second important feature, low holding values, representing negative holding positions due to short selling, contribute to the ``buy'' action. When the 60-min momentum shows an upward trend, the agent prefers to place ``buy'' orders. Therefore, we can infer that the agent trained without bubble experience learned a ``momentum-based'' trading strategy. The agent follows the emerging short-term trends in the order book and places orders to intensify the trends. Understanding that the market volatility of the scenarios without a bubble is relatively low compared to the bubble scenarios, such a short-term momentum-based strategy can benefit the agent by following the stock's fundamental value and making profits. However, in bubble scenarios, closely tracing the short-term momentum and placing orders to aggravate the order book imbalance will possibly cause huge losses.

The agent trained with 100\% bubble experience also relies on the ``imbalance'' feature, ranked as the second feature shown in Figure \ref{fig_shap_mkt_100}. However, according to the SHAP value shown in the horizontal axis, the general influence of ``imbalance'' was lower compared to the other agent. High imbalance values, indicating overwhelmed ``buy'' orders in the order book, contributed less to placing buy orders than the agent without bubble experience. Also, long-term price momentum features, including 120 and 180-min, majorly contributed to the decisions. Unlike short-term momentum tracing all price trends, 120 and 180-min momentum filter price surges caused by minor variances and capture significant price moves with a longer duration in bubble periods. In this scenario, high long-term momentum values led to placing ``sell'' orders that suppressed bubbles. This reveals that the trading strategy of this agent majorly relies on the stock's intrinsic value and is more conservative to the imbalanced order book and price surges.

While training the RL agents, momentum and herding agents will push the price towards a certain direction and create large volatility to form bubbles. However, such a direction can be upward or downward due to the randomness in the fundamental value generated by an O-U process. So, the agent trained with 100\% bubble experience would experience large market volatility in two possible directions. Such uncertainty and large volatility in the market cause the agent to learn conservative trading strategies, which rely on long-term momentum, and place orders to balance the order book.

Therefore, we can conclude that these two agents demonstrated different trading strategies. The agent trained without bubble experience closely follows the short-term price momentum and tends to place orders to the dominant side in the order book to intensify existing momentum, thereby acting like momentum traders. In contrast, the other agent learned a more conservative trading strategy, in which the agent monitored long-term momentum and placed orders to suppress significant price surges back towards the fundamental values, acting like value traders. Such strategy differences led to a considerable difference in profit and bubble measurements.

\section{Conclusion}
\label{ch_conclusion}

In this work, we investigated the impact of experiencing asset price bubbles during training on the converged trading strategies of reinforcement learning traders. We expected that RL agents trained with a higher percentage of bubble scenarios would better suppress future bubbles. We specified three bubble measurement metrics: count, magnitude, and duration. RL agents were trained with different percentages (0\%, 25\%, 50\%, 75\%, and 100\%) of bubble scenarios and applied to test sessions containing at least one bubble. Experimental results show significant differences in agent performance across all bubble measurements and support our hypothesis that the agent's ability to suppress bubbles increases with the percentage of bubble scenarios in the agent's training.

In this study, market bubbles burst via a regulatory time constraint. In future work, we plan to investigate bubbles burst by liquidity shortage and examine how this affects market equilibrium when bubbles are present. Also, we used a fixed order size for the actions of the RL agents. Since these agents only act once per minute, they cannot quickly accumulate a large position. In future research, we will implement a continuous action space for more dynamic order sizes. As the next step, we will use historical market bubble prices to calibrate our bubble generation mechanism and further evaluate our trained agents to measure their profit capabilities and performance of suppressing bubbles. In summary, learning from experience can enhance RL traders' ability to handle unusual market conditions like bubbles. Learning agents exposed to bubbles in training can reduce market volatility and potentially prevent bubble occurrence. The results from this study provide us insight that exposure to rare and extreme market scenarios is important for the safety of developing large-scale autonomous trading systems.

\section*{ACKNOWLEDGMENTS}

This paper was prepared for informational purposes in part by the Artificial Intelligence Research group of JPMorgan Chase \& Co and its affiliates (“J.P. Morgan”) and is not a product of the Research Department of J.P. Morgan. J.P. Morgan makes no representation and warranty whatsoever and disclaims all liability for the completeness, accuracy, or reliability of the information contained herein. This document is not intended as investment research or advice, or a recommendation, offer, or solicitation for the purchase or sale of any security, financial instrument, financial product, or service, or to be used in any way for evaluating the merits of participating in any transaction, and shall not constitute a solicitation under any jurisdiction or to any person if such solicitation under such jurisdiction or to such person would be unlawful.

\footnotesize

\bibliographystyle{wsc}

\bibliography{wsc23paper}

\section*{AUTHOR BIOGRAPHIES}


\noindent {\bf HAIBEI ZHU} is a Research Scientist at J.P. Morgan AI Research, concentrating on synthetic data, multi-agent systems, and reinforcement learning. He received his Ph.D. from Duke, where his research centered on human-computer interaction and machine learning. Upon joining AI Research, he focused on simulating market scenarios with financial bubbles within multi-agent environments and training reinforcement learning traders. He is also engaged in research dedicated to generating high-fidelity synthetic time series data. His email address is \email{haibei.zhu@jpmchase.com}.\\

\noindent {\bf SVITLANA VYETRENKO} is a Research Director at J.P. Morgan AI Research and a Lecturer at UC Berkeley. She holds a Ph.D. from the California Institute of Technology, where she also earned her Master of Science. Svitlana has over a decade of experience in the financial industry. Her expertise spans quantitative research, machine learning, multi-agent simulations, and reinforcement learning. As a Research Director, she leads projects on market simulations and time series synthetic data. With a passion for cutting-edge research, Svitlana is dedicated to advancing the field of artificial intelligence and its applications in the financial industry. Her email address is \email{svitlana.s.vyetrenko@jpmchase.com}.\\

\noindent {\bf SERAFIN GRUNDL} is a Principal Economist at the Federal Reserve Board of Governors in the Division for Research \& Statistics. His research is mainly focused on industrial organization; in particular, identification and estimation of first-price auctions, and various topics in antitrust (e.g. mergers \& common ownership). In his policy work, he assesses the competitive effects of bank mergers and other antitrust issues in the banking industry. His email address is \email{serafin.j.grundl@frb.gov}.\\

\noindent {\bf DAVID BYRD} is an Assistant Professor of Computer Science at Bowdoin College, where he teaches courses in artificial intelligence and financial machine learning. His current areas of research interest include normative guidance for intelligent financial agents, market simulation, automated portfolio inference, and deep reinforcement learning. Dr. Byrd holds a Ph.D. in computer science from Georgia Tech, for which he developed the ABIDES simulation engine, now widely used to study complex systems of distributed intelligent agents. His most recent work investigates how intelligent agents may inadvertently learn to spoof the stock market, and how we might discourage them. His email address is \email{d.byrd@bowdoin.edu}.\\

\noindent {\bf KSHAMA DWARAKANATH} is a Research Scientist at J.P. Morgan AI Research working on using reinforcement learning to design and learn trading agents with diverse objectives in simulated multi-agent markets. She completed her Master’s degree at UC Berkeley, focused on nonlinear systems and control engineering. Her interests lie in the fields of reinforcement learning, multi-agent simulations, and mechanism design. Her email address is \email{kshama.dwarakanath@jpmorgan.com}.\\

\noindent {\bf TUCKER BALCH} is a computer scientist, researcher, and educator specializing in AI, Robotics, and Finance. He earned his Bachelor's degree and Ph.D. in Computer Science from the Georgia Institute of Technology. Dr. Balch has published over 120 peer-reviewed research papers. He has held research and teaching positions at Carnegie Mellon University and Georgia Institute of Technology. In the financial industry, Dr. Balch co-founded Lucena Research. He currently serves as a managing director at J.P. Morgan AI Research, leading teams exploring ML and Cryptography, multi-agent system simulations, high-frequency electronic markets, and Synthetic Data for Finance. He is also an online education pioneer, developing and teaching MOOCs like "Computational Investing" and "Machine Learning for Trading," which have reached over 170,000 students worldwide. Before his research career, Balch served as an F-15 pilot in the U.S. Air Force. His email address is \email{tucker.balch@jpmchase.com}.

\end{document}

%% file: wscbib.tex
\makeatletter
\let\@internalcite\cite
\def\cite{\def\@citeseppen{-1000}%
    \def\@cite##1##2{(##1\if@tempswa , ##2\fi)}%
    \def\citeauthoryear##1##2##3{##1 ##3}\@internalcite}
\def\citeNP{\def\@citeseppen{-1000}%
    \def\@cite##1##2{##1\if@tempswa , ##2\fi}%
    \def\citeauthoryear##1##2##3{##1 ##3}\@internalcite}
\def\citeN{\def\@citeseppen{-1000}%
    \def\@cite##1##2{##1\if@tempswa, ##2)\else{}\fi}%
    \def\citeauthoryear##1##2##3{##1 (##3)}\@citedata}
\def\citeA{\def\@citeseppen{-1000}%
    \def\@cite##1##2{(##1\if@tempswa , ##2\fi)}%
    \def\citeauthoryear##1##2##3{##1}\@internalcite}
\def\citeANP{\def\@citeseppen{-1000}%
    \def\@cite##1##2{##1\if@tempswa , ##2\fi}%
    \def\citeauthoryear##1##2##3{##1}\@internalcite}
\def\shortcite{\def\@citeseppen{-1000}%
    \def\@cite##1##2{(##1\if@tempswa , ##2\fi)}%
    \def\citeauthoryear##1##2##3{##2 ##3}\@internalcite}
\def\shortciteNP{\def\@citeseppen{-1000}%
    \def\@cite##1##2{##1\if@tempswa , ##2\fi}%
    \def\citeauthoryear##1##2##3{##2 ##3}\@internalcite}
\def\shortciteN{\def\@citeseppen{-1000}%
    \def\@cite##1##2{##1\if@tempswa, ##2\else{}\fi}%
    \def\citeauthoryear##1##2##3{##2 (##3)}\@citedata}
\def\shortciteA{\def\@citeseppen{-1000}%
    \def\@cite##1##2{(##1\if@tempswa , ##2\fi)}%
    \def\citeauthoryear##1##2##3{##2}\@internalcite}
\def\shortciteANP{\def\@citeseppen{-1000}%
    \def\@cite##1##2{##1\if@tempswa , ##2\fi}%
    \def\citeauthoryear##1##2##3{##2}\@internalcite}
\def\citeyear{\def\@citeseppen{-1000}%
    \def\@cite##1##2{(##1\if@tempswa , ##2\fi)}%
    \def\citeauthoryear##1##2##3{##3}\@citedata}
\def\citeyearNP{\def\@citeseppen{-1000}%
    \def\@cite##1##2{##1\if@tempswa , ##2\fi}%
    \def\citeauthoryear##1##2##3{##3}\@citedata}
%
%
%
\def\@citedata{%
    \@ifnextchar [{\@tempswatrue\@citedatax}%
                  {\@tempswafalse\@citedatax[]}%
}

\def\@citedatax[#1]#2{%
\if@filesw\immediate\write\@auxout{\string\citation{#2}}\fi%
  \def\@citea{}\@cite{\@for\@citeb:=#2\do%
    {\@citea\def\@citea{, }\@ifundefined
       {b@\@citeb}{{\bf ?}%
       \@warning{Citation `\@citeb' on page \thepage \space undefined}}%
{\csname b@\@citeb\endcsname}}}{#1}}%

%
\def\@citex[#1]#2{%
\if@filesw\immediate\write\@auxout{\string\citation{#2}}\fi%
  \def\@citea{}\@cite{\@for\@citeb:=#2\do%
    {\@citea\def\@citea{; }\@ifundefined
       {b@\@citeb}{{\bf ?}%
       \@warning{Citation `\@citeb' on page \thepage \space undefined}}%
{\csname b@\@citeb\endcsname}}}{#1}}%

%
\def\@biblabel#1{}
\makeatother



\newdimen\bibindent
\bibindent=0.0em
\def\thebibliography#1{\section*{\refname}\list
   {}{\settowidth\labelwidth{[#1]}
   \leftmargin\parindent
   \itemindent -\parindent
   \listparindent \itemindent
   \itemsep 0pt
   \parsep 0pt}
   \def\newblock{}
   \sloppy
   \sfcode`\.=1000\relax}